\documentclass[prd]{revtex4}
\usepackage{amsmath,amssymb,latexsym}
\usepackage{graphicx}

\def\la{\lambda}
\def\La{\Lambda}
\def\ve{\varepsilon}
\def\<{\langle}
\def\>{\rangle}
\def\GeV{{\rm\ GeV}}
\def\up{\uparrow}
\def\down{\downarrow}

\def\Re{\mathop{\rm Re}\nolimits}

\def\CG{{\cal G}}
\def\CM{{\cal M}}

\begin{document}
\title{Perturbative QCD predictions for two-photon exchange}
\author{Dmitry~Borisyuk}
\author{Alexander~Kobushkin}

\affiliation{Bogolyubov Institute for Theoretical Physics\\
Metrologicheskaya street 14-B, 03680, Kiev, Ukraine}

%\maketitle

\begin{abstract}
We study two-photon exchange (TPE) in the elastic electron-nucleon scattering
at high $Q^2$ in the framework of pQCD.
The obtained TPE amplitude is of order $\alpha/\alpha_s$
with respect to Born approximation.
Its shape and value are sensitive to the choice of nucleon wavefunction,
thus study of TPE effects can provide important information about nucleon structure.
With the wavefunctions based on QCD sum rules,
TPE correction to the electron-proton cross section has negative sign,
is almost linear in $\ve$ and grows logarithmically with $Q^2$
up to 7\% at $Q^2 = 30\GeV^2$. % and low $\ve$.
%At $Q^2 = 30\GeV^2$ and low $\ve$ it reaches 7\%.
The results of existing "hadronic" calculations,
taking into account just the nucleon intermediate state,
can be smoothly connected with pQCD result near $Q^2 \sim 3\GeV^2$.
Above this point two methods disagree,
which implies that "hadronic" approach becomes inadequate at high $Q^2$.
Other relevant observables, such as electron/positron
cross section ratio, are also discussed.
\end{abstract}

\maketitle 

\section{Introduction}
 The two-photon exchange (TPE) in electron-proton scattering
 is actively discussed over last several years.
 The impetus for this was initially given by the discovery of so-called
 $G_E/G_M$ problem in the proton form factor measurements \cite{GE/GM}.
 It was shown later that the discrepancy between Rosenbluth separation
 and polarization transfer methods can be at least partially
 eliminated after taking into account TPE effects \cite{BMT}.
 Several experiments aimed
 at direct detection of TPE contribution to cross section,
 are proposed \cite{expts}.
 Non-zero single-spin asymmetry, which is induced by the imaginary part
 of TPE amplitude, was observed experimentally \cite{SSA}.
 The role of TPE in determination of proton radius \cite{radius},
 parity-violating observables \cite{parity} and 
 in deep inelastic scattering \cite{DIS} was also discussed.
 % Further review and a bibliography may be found in Ref.~\cite{Rev}.
 
 Currently, the measurements of proton form factors at $Q^2 \sim 10 \GeV^2$
 are in progress \cite{GEpIII}
 and other measurements in this region are proposed \cite{highQ2-p,highQ2-n}.
 Clearly, these experiments call
 for the reliable estimate of TPE effects for high-$Q^2$ kinematics,
 which was one of the aims of the present work.
 At moderate $Q^2$, the TPE amplitude was calculated using nucleon and
 resonances as intermediate states (further called "hadronic" approach)
 \cite{BMT,OurBox,BMTRes,OurDisp}.
 At high $Q^2$, however, a natural means for the description of
 any process involving hadrons, and in particular TPE,
 is perturbative  quantum chromodynamics (pQCD).
 Surprisingly, we have found no direct pQCD calculation
 of TPE in the literature.
 
 In Ref.~\cite{Parton}, TPE at high $Q^2$ was investigated using the formalism
 of generalized parton distributions.
 The authors doubt of pQCD applicability
 in the presently accessible kinematical region and
 thus use an alternative method.
 The values of TPE corrections obtained this way have opposite sign
 to the results of "hadronic" calculations.
 The authors also use an assumption that the most important
 diagrams are that in which both photons interact with the same quark.
 It turns out that in pQCD approach the situation is reversed
 (see below, Sec.~\ref{sec:TPE}).
 
 In the present paper we study TPE in the elastic electron-nucleon scattering
 at high $Q^2$ in the framework of pQCD. We employ the method,
 which was used to calculate baryon form factors in Refs.~\cite{BL,CZ}.
 In the adopted approach, a nucleon with momentum $p$ is represented
 as three collinearly moving quarks with momenta $x_i p$,
 where $0<x_i<1$, $\sum_{i=1}^3 x_i = 1$.
 All quark masses and nucleon mass are neglected and thus $p^2=(x_i p)^2=0$.
 The process amplitude has the form
\begin{equation} \label{fTf}
 \CM = \<\phi(y_i) | T(y_i,x_i) | \phi(x_i) \>
\end{equation}
where $T$ is hard scattering amplitude at quark level
(represented by appropriate Feynman diagrams), 
$\phi(x_i)$ and $\phi(y_i)$ are initial and final nucleon
spin-flavor-coordinate wavefunctions (quark distribution amplitudes).
The convolution with nucleon wavefunction implies
a convolution of spinor indices and an integration over
$dx_1 dx_2 dx_3 \delta(1-x_1-x_2-x_3)$ and similarly for $y_i$.

To obtain non-zero transition amplitude one must turn
 the momenta of all three quarks from initial to final direction.
 In one-photon exchange (Born) approximation
 one therefore needs at least two hard gluons to be exchanged between the quarks.
 It follows then that the amplitude scales as ${\alpha\alpha_s^2}/{Q^6}$
 and the nucleon form factor --- as ${\alpha_s^2}/{Q^4}$.
 In the case of TPE the exchange of one gluon is sufficient and thus
 the leading-order pQCD contribution to the TPE amplitude should be
 $\sim {\alpha^2\alpha_s}/{Q^6}$.
 The ratio TPE/Born then will be not just $\alpha$, as one may naively expect,
 but $\alpha/\alpha_s$, which is significantly larger and growing with $Q^2$.
 Thus the larger $Q^2$ is, the more important TPE will be.

 The paper is organized as follows. In Sec.~\ref{sec:kin} the observables,
 affected by TPE, are discussed, Sec.~\ref{sec:ampl} describes all ingredients
 of calculation (hard scattering amplitude for one-photon and
 two-photon exchange and nucleon wavefunctions),
 numerical results are given in Sec.~\ref{sec:res}
 and conclusions --- in Sec.~\ref{sec:concl}.
 
\section{Kinematics and observables} \label{sec:kin}

 The momenta of particles are denoted according to
 $e(k)+N(p) \to e(k')+N(p')$. The transferred momentum is $q = p'-p'$,
 $Q^2 = -q^2 > 0$ and $\nu = (p+k)(p'+k')$.
 The reduced cross section of elastic electron-proton scattering
 can be written as
\begin{equation} \label{++}
  \sigma_R = \frac{Q^2}{4M^2} |\CG_M|^2 + \ve |\CG_E|^2 +
  \frac{Q^2}{4M^2} \frac{\ve(1-\ve)}{1+\ve} |\CG_3|^2
\end{equation}
 where $M$ is proton mass, $\ve = 1 - 2[1+\nu^2/Q^2(4M^2+Q^2)]^{-1}$,
 which for $Q^2 \gg M^2$ turns to
\begin{equation} \label{epsilon}
 \ve \approx (\nu^2-Q^4)/(\nu^2+Q^4)
\end{equation}
 and $\CG_M$, $\CG_E$ and $\CG_3$ are
 certain invariant amplitudes (see details in Ref.\cite{OurDisp}).
 In Born approximation $\CG_E$ and $\CG_M$ become usual
 electric and magnetic form factors and $\CG_3$ vanishes, that is
\begin{equation}
 \CG_E = G_E + \delta\CG_E, \qquad
 \CG_M = G_M + \delta\CG_M, \qquad
 \CG_3 = \delta\CG_3,
\end{equation}
 where the prefix $\delta$ indicates TPE contribution.
 The dominant part of the cross section at high $Q^2$
 comes from the generalized magnetic form factor $\CG_M$
 and can be written as
 \begin{equation} \label{sigma}
  \sigma_R \approx \frac{Q^2}{4M^2} G_M^2 \left(1 + 2\Re\frac{\delta\CG_M}{G_M} \right)
 \end{equation}
 Hence we should primarily study TPE contribution $\delta\CG_M$.
 This quantity also defines the positron/electron
 elastic cross section ratio:
\begin{equation} \label{R+-}
 R = \frac{\sigma^+}{\sigma^-} =
  \left| \frac{G_M-\delta\CG_M}{G_M+\delta\CG_M} \right|^2 \approx
  1 - 4 \Re \frac{\delta\CG_M}{G_M}
\end{equation}

 The generalized electric form factor $\CG_E$ is suppressed
 in the cross section by $M^2/Q^2 \ll 1$.
 Therefore a consideration of TPE corrections
 to $\CG_E$ makes little sense.
 
 The amplitude $\delta\CG_3$ in principle can be measured
 in polarization experiments.
 Namely, neglecting terms of order $M^2/Q^2$, the polarization of final proton
 in recoil polarization method is purely longitudinal and equals
\begin{equation} \label{P_l}
 P_\ell = h \sqrt{1-\ve^2}
  \left( 1 - \frac{2\ve^2}{1+\ve} \Re \frac{\delta\CG_3}{G_M} \right)
\end{equation}
 where $h$ is electron helicity. Thus a precise study of
 $\ve$-dependence of $P_\ell$ may give an access to $\delta\CG_3$.

\section{Amplitude calculation} \label{sec:ampl}

\subsection{One-photon exchange}
At first, we briefly review the pQCD calculation of form factors
\cite{BL,CZ}.
\begin{figure}%[b]
\def\www{0.13}
 \centering
 (a)
 \includegraphics[width=\www\textwidth]{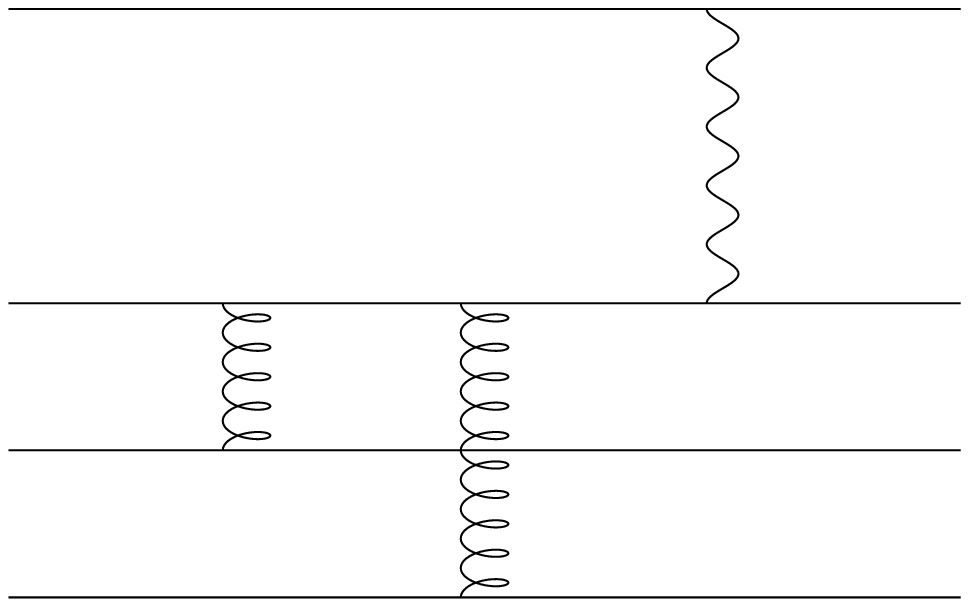}\hfil
 \includegraphics[width=\www\textwidth]{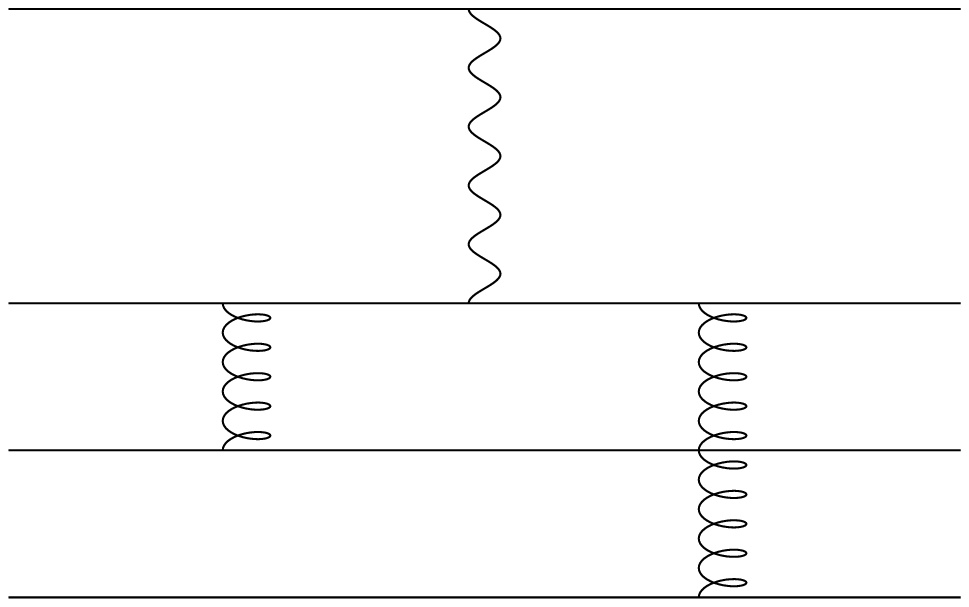}\hfil
 \includegraphics[width=\www\textwidth]{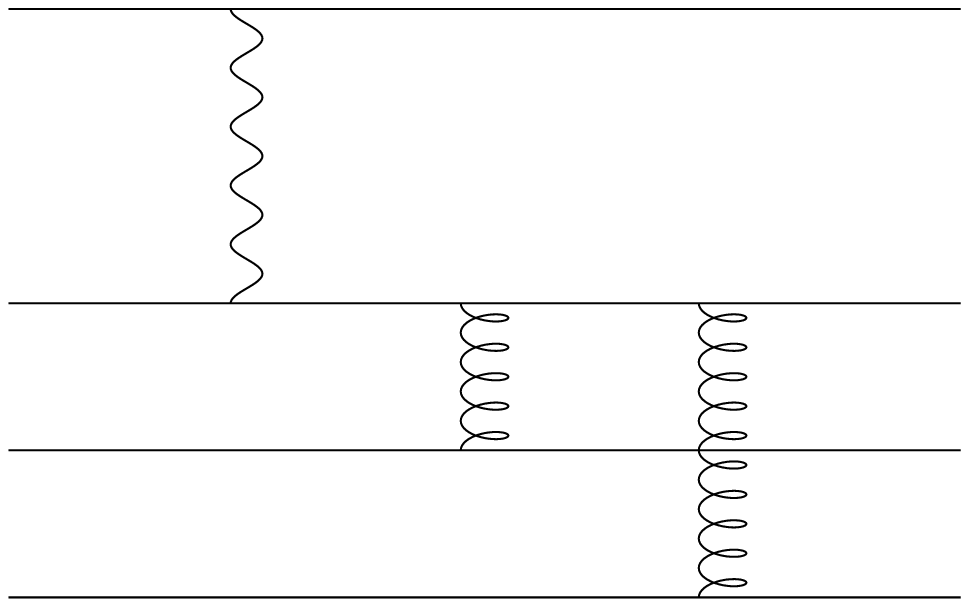}\hfil
 \includegraphics[width=\www\textwidth]{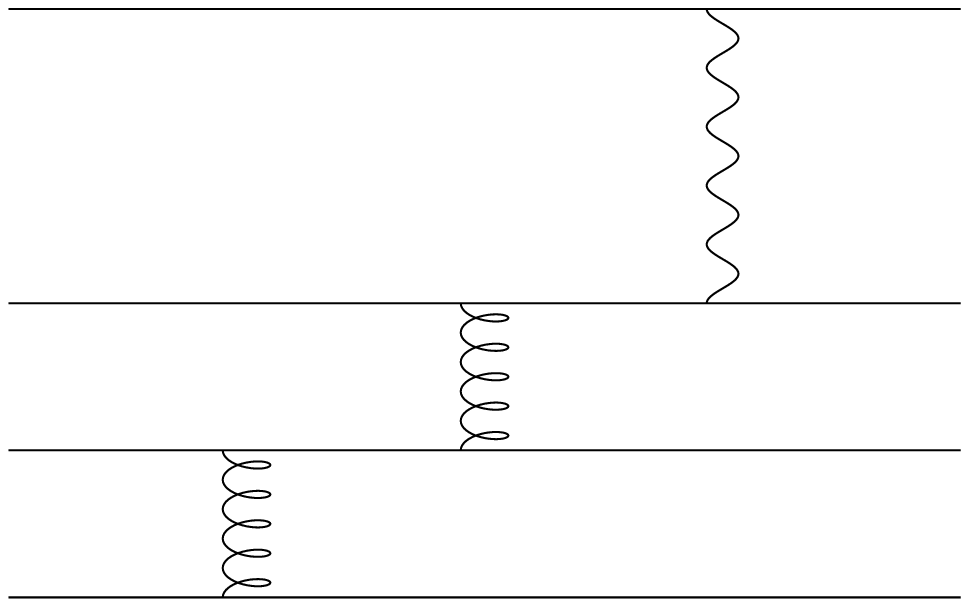}\hfil
 \includegraphics[width=\www\textwidth]{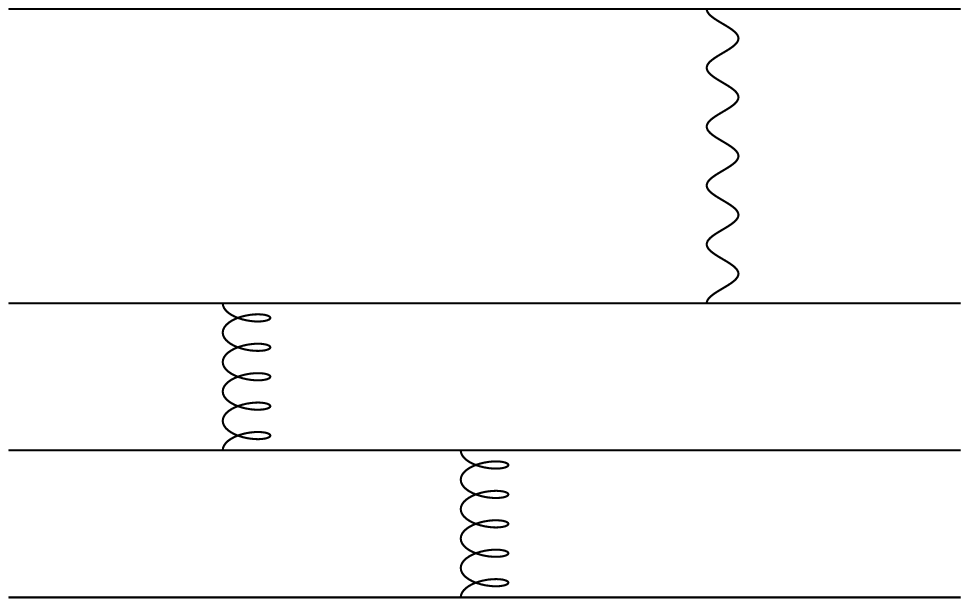}\hfil
 \includegraphics[width=\www\textwidth]{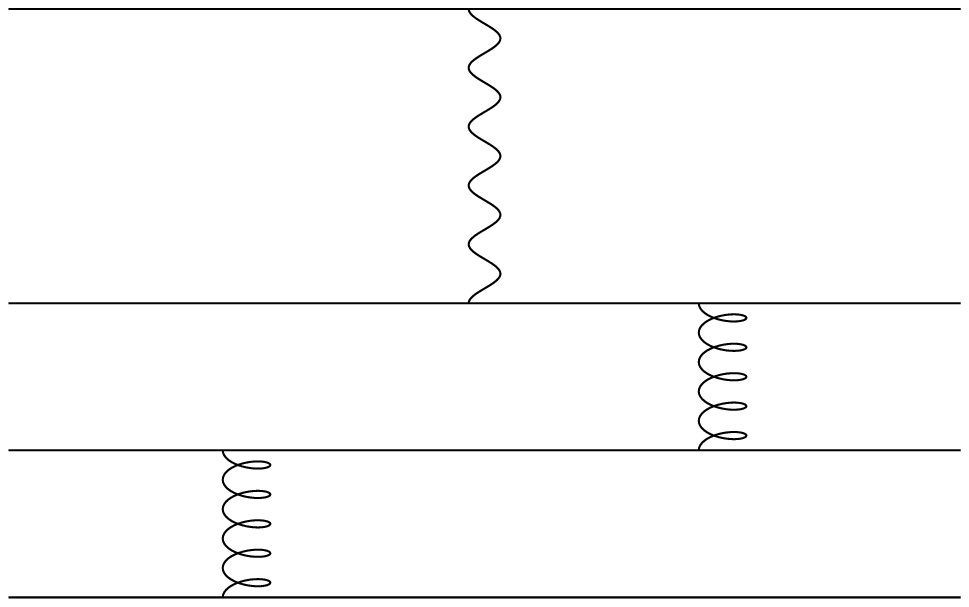}\hfil
 \includegraphics[width=\www\textwidth]{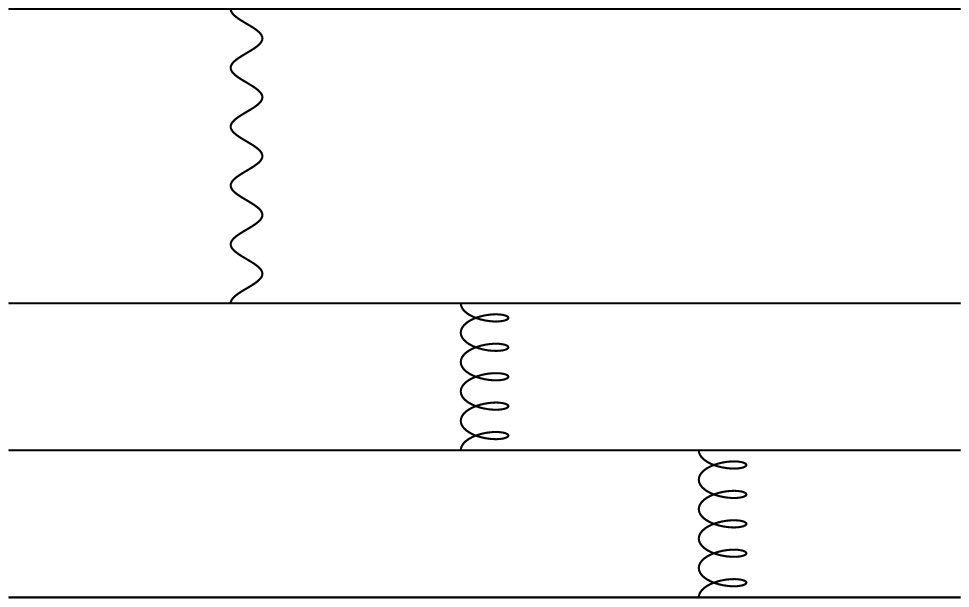}\\[5mm]
 (b)
 \includegraphics[width=\www\textwidth]{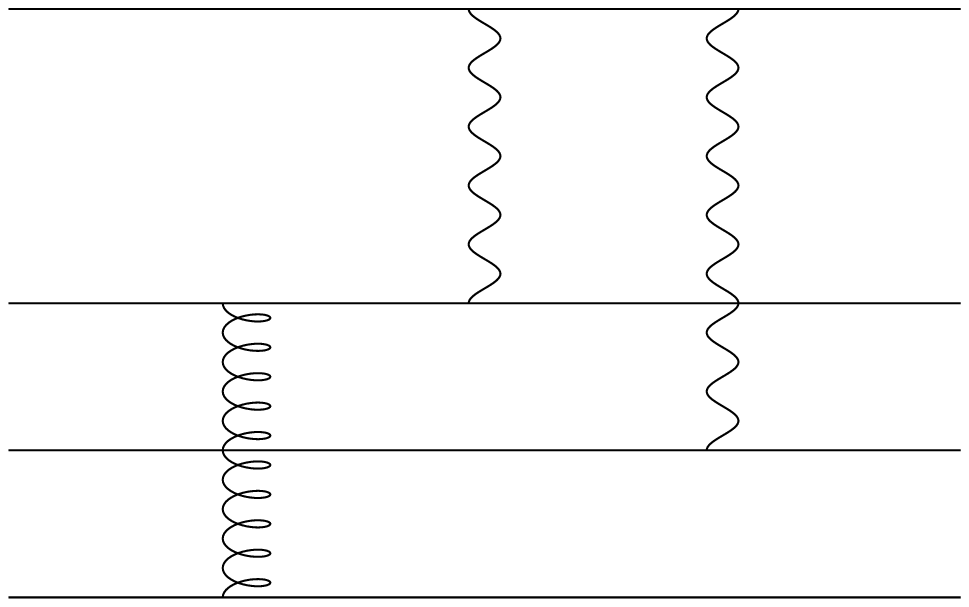}\hfil
 \includegraphics[width=\www\textwidth]{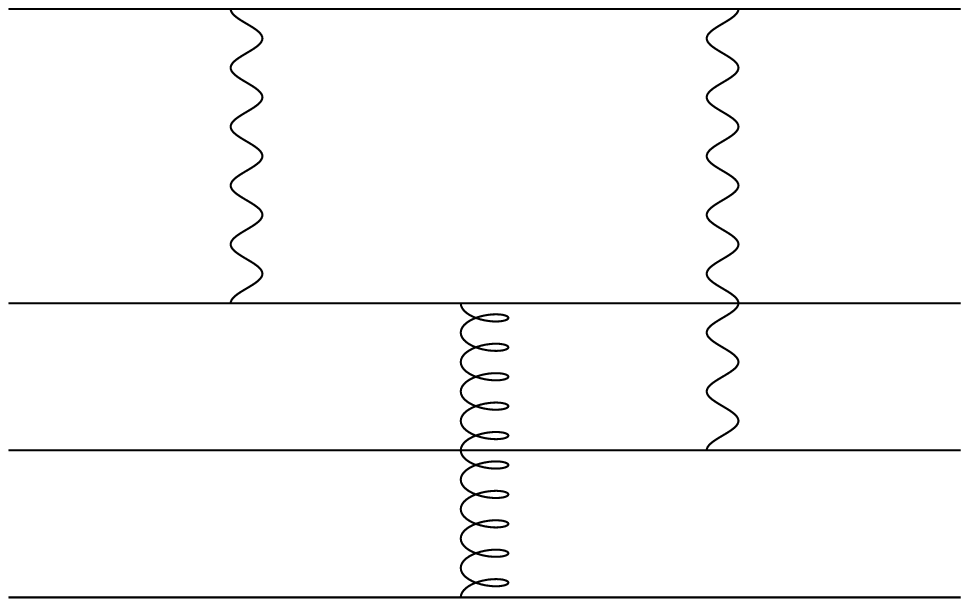}\hfil
 \includegraphics[width=\www\textwidth]{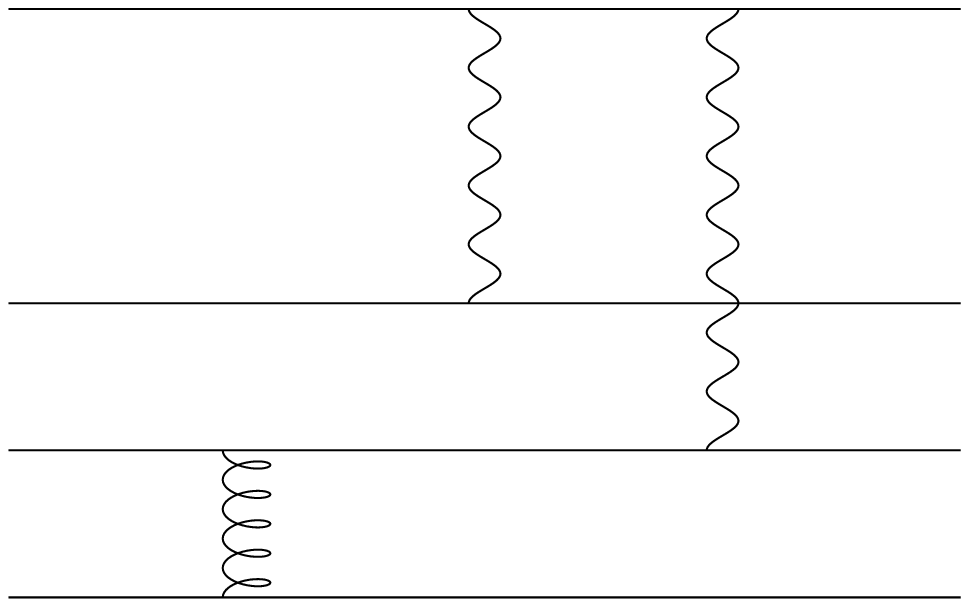}\hfil  
 \includegraphics[width=\www\textwidth]{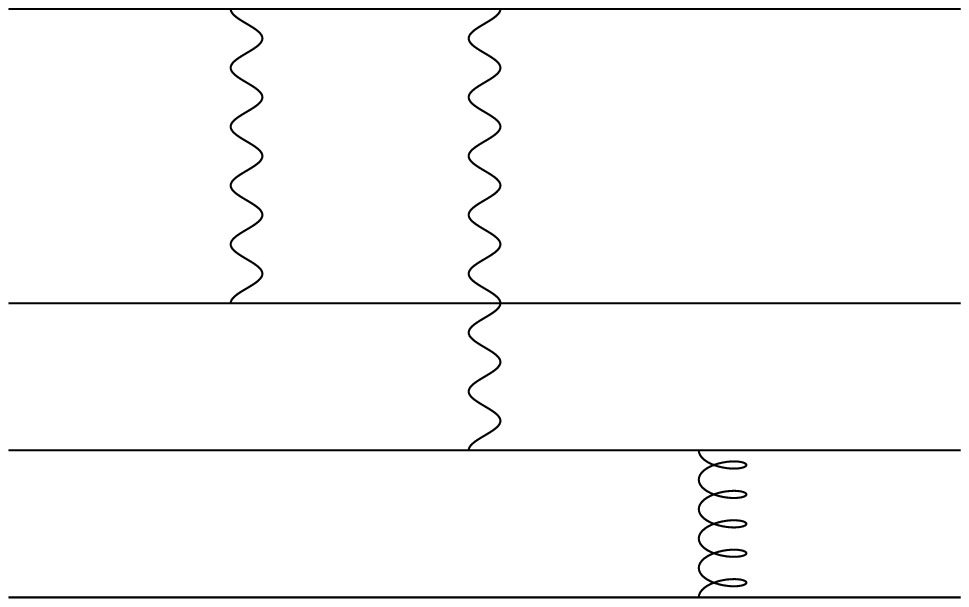}\hfil
 (c)
 \includegraphics[width=\www\textwidth]{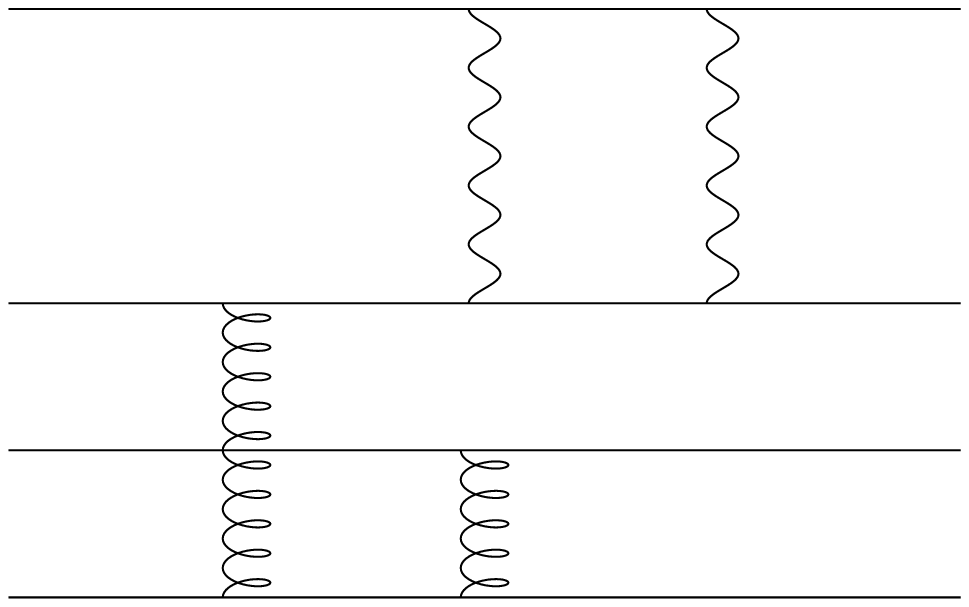}\hfil
 (d)
 \includegraphics[width=\www\textwidth]{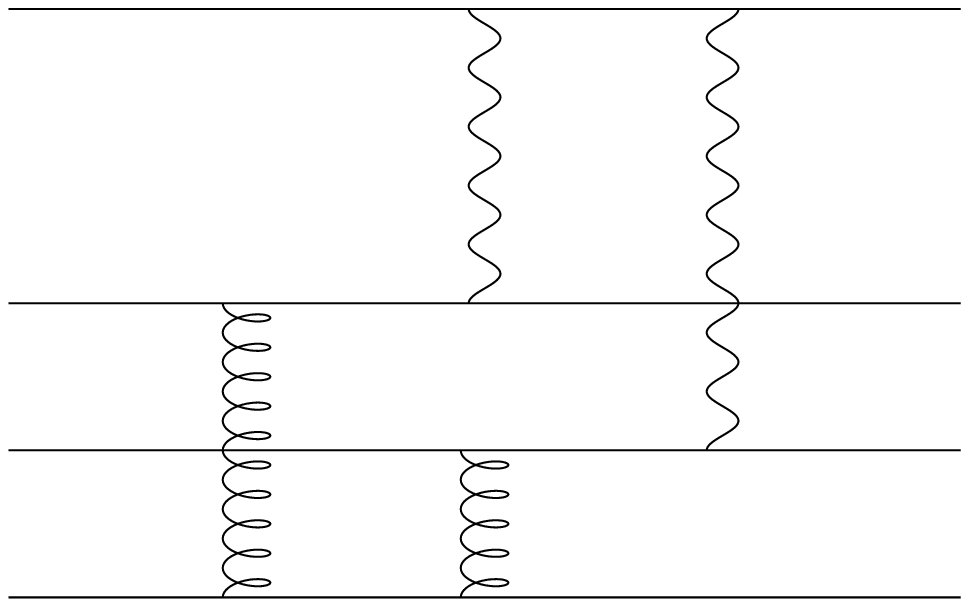}
 \caption{pQCD diagrams for $eN\to eN$: one-photon exchange (a),
 two-photon exchange, leading order (b), subleading order~(c,d).}
 \label{diagr}
\end{figure}
 The nucleon form factors, or the elastic electron-nucleon scattering
 in one-photon exchange approximation, is described in pQCD
 to leading order in $\alpha_s$ by 7 diagrams (Fig.~\ref{diagr}a).
 For example, the piece of amplitude, coming from the first of them, is
\begin{eqnarray} \label{*}
 \CM & = & - \frac{4\pi\alpha}{q^2} \left(\frac{4\pi\alpha_s}{q^2}\right)^2
  \cdot 6 \cdot ( 4/9 ) \cdot (-q^2) \cdot \bar u'\gamma_\mu u \times \\
 && \< \phi(y_i) | e_1 \frac{
     \gamma_\alpha (y_1\hat p'+y_2\hat p'-x_2\hat p) \gamma_\beta
     (x_1\hat p +\hat q) \gamma_\mu \otimes \gamma_\alpha \otimes \gamma_\beta}
    {x_2^2 y_2 x_3 y_3 (y_1+y_2)(1-x_1) q^4} | \phi(x_i) \> \nonumber
\end{eqnarray}
 where $u$ and $u'$ are electron spinors, 
 the overall minus sign is due to $negative$ electron charge,
 $e_i$ is $i$th quark charge,
 $6=3!$ is a symmetry factor due to possible permutations of quark lines,
 $(-q^2)$ comes from quark and nucleon spinors normalization,
\begin{equation}
 4/9 = \<\tfrac12\la^a \tfrac12\la^b \otimes \tfrac12\la^a \otimes \tfrac12\la^b\>
\end{equation}
 is a color factor (here $\la^a$ are Gell-Mann matrices and $\<\ \>$ means
 averaging over totally antisymmetric color wavefunction).
 In the last equation as well as in Eq.~(\ref{*}) we separate matrices,
 acting on different quarks, by a $\otimes$ sign.
 Thus in the expression for color factor $\frac12\la^a \frac12\la^b$
 acts on the first, $\frac12\la^a$ on the second and $\frac12\la^b$
 on the third quark.
 
 Adding up contributions from all diagrams and using the fact that
 spin-flavor-coordinate wavefunction is totally symmetric
 under quark interchange, we obtain
\begin{equation}
 \CM = - \frac{4\pi\alpha}{q^2} \bar u'\gamma_\mu u \cdot
 \bar U'\gamma_\mu U \cdot G_M(q^2)
\end{equation}
 where $U$ and $U'$ are initial and final nucleon spinors and
\begin{eqnarray} \label{**}
 G_M & = & \frac{16}{3} \left( \frac{4\pi\alpha_s}{q^2} \right)^2
  \< \phi(y_i) | (1 + h_1 h_3) \left\{
     \frac{2e_1}{x_3 y_3 (1-x_1)^2 (1-y_1)^2} + 
     \frac{2e_1}{x_2 y_2 (1-x_1)^2 (1-y_1)^2} + 
     \right. \\ && \left.     
     \frac{e_2}{x_1 y_1 x_3 y_3 (1-x_1) (1-y_3)} -
     \frac{e_1}{x_2 y_2 x_3 y_3 (1-x_1) (1-y_3)} -
     \frac{e_1}{x_2 y_2 x_3 y_3 (1-x_3) (1-y_1)}
  \right\} | \phi(x_i) \> \nonumber
\end{eqnarray}
 where $h_i=\pm 1$ are signs of quark helicities;
 the helicities of initial and final quarks should be equal.
 This is equivalent to well-known result \cite{BL,CZ}.

\subsection{Two-photon exchange} \label{sec:TPE}
 For the case of TPE, there are only 4 distinct diagrams in the leading order
 (Fig.~\ref{diagr}b), in which photons are connected to different quarks.
 The diagrams in which both photons interact with the same quark
 (Fig.~\ref{diagr}c), need one more gluon to turn all quarks' momenta
 and thus are subleading in $\alpha_s$.
 Moreover, the evaluation of such diagrams alone is inconsistent,
 since the contribution of the same order in $\alpha_s$ comes from
 one-gluon corrections to the leading diagrams (e.g. Fig.~\ref{diagr}d).
 
 One point needs to be clarified here. If we remove the electron line,
 the diagrams Fig.~\ref{diagr}b-d will represent Compton scattering
 of virtual photons on the nucleon (doubly virtual Compton scattering, VVCS).
 And vice versa, TPE can be viewed as a process in which
 the virtual photon, emitted by the electron,
 is scattered from the proton and then absorbed back by the electron.
 VVCS has an important qualitative difference from the well-studied
 {\it real} Compton scattering (RCS). Since the momentum $r$ of
 the real photon satisfies $r^2=0$, it cannot alone turn quark's momentum:
 $ (x_i p - y_i p')^2 \ne 0 = r^2$.
 Therefore diagrams with the structure like (Fig.~\ref{diagr}b) vanish for RCS,
 and the amplitude expansion begins with $O(\alpha_s^2)$ terms
 (diagrams like Fig.~\ref{diagr}c,d). On contrary, VVCS photons
 may be highly virtual, diagrams Fig.~\ref{diagr}b contribute,
 and leading terms in VVCS amplitude are $O(\alpha_s)$.
 Hence one cannot employ an analogy with RCS in the analysis of TPE
 (cf. Ref.~\cite{Parton}).
 
 We write down the expression for the first diagram in Fig.~\ref{diagr}b,
 the rest are analogous. We have
\begin{eqnarray}
 \delta\CM & = & \left(\frac{4\pi\alpha}{q^2}\right)^2 \frac{4\pi\alpha_s}{q^2} \cdot
 6 \cdot \left(-2/3 \right) \cdot (-q^2) \times \\
 && \< \phi(y_i) | e_1 e_2
       \frac{\bar u' \gamma_\mu (\hat k + x_2 \hat p - y_2 \hat p') \gamma_\nu u}
         {(k + x_2 p - y_2 p')^2 +i0}
       \cdot
       \frac{\gamma_\alpha(y_1\hat p'+y_3\hat p'-x_3\hat p)\gamma_\mu
          \otimes \gamma_\nu \otimes \gamma_\alpha }
          {x_2 y_2 x_3^2 y_3 (x_1+x_3) (y_1+y_3)^2 q^2}
 | \phi(x_i) \> \nonumber
\end{eqnarray}
 where the color factor is
 $-2/3 = \< \frac12\la^a \otimes \frac12\la^a \otimes 1\>$.
 After some algebraic transformations and using wavefunction symmetry,
 we obtain the full TPE amplitude in the form
\begin{equation} \label{x}
 \delta\CM = - \frac{4\pi\alpha}{q^2} \bar u'\gamma_\mu u \cdot
   \bar U' \gamma_\nu U \cdot
   \left( \frac{4 p_\mu k_\nu}{\nu} \delta\CG_3 + g_{\mu\nu} \delta G_M \right)
\end{equation}
 where
\begin{eqnarray} \label{delta}
 (\delta G_M,\delta\CG_3) & = &
 -\frac{256\pi^2\alpha\alpha_s}{q^4} \< \phi(y_i) |
   \frac{e_1 e_2 (1 - h_1 h_3)}{x_2 y_2 x_3 y_3 (1-x_2)(1-y_2)} \times \\
 && \frac{1}{\nu(x_2-y_2)-q^2(x_2+y_2-2 x_2 y_2)+i0}
    \left( \frac{\nu-q^2}{1-x_2} + \frac{\nu+q^2}{1-y_2} - 2\nu, 2\nu \right)
%    \frac{1}{2 x_2 pk - 2 y_2 p'k + x_2 y_2 q^2 + i0}
%    \left( \frac{x_2 pk}{1-x_2} + \frac{y_2 p'k}{1-y_2}, \nu \right)
   | \phi(x_i) \> \nonumber
\end{eqnarray}
 From this expression it is easy to see the crossing symmetry of TPE
 amplitudes $\delta G_M$ and $\delta\CG_3$: both are $\nu$-odd.

 The quantity $\delta\CG_M$, associated with the cross section correction
 (\ref{sigma}), equals \cite{OurDisp}
\begin{equation}
 \delta\CG_M = \delta G_M + \ve \delta\CG_3
\end{equation}
 As implied by Eqs.~(\ref{sigma}-\ref{P_l}), it is better to consider ratios
 $\delta\CG_M/G_M$ and $\delta\CG_3/G_M$ than the amplitudes themselves.
 This way we also avoid the uncertainty related with
 the absolute normalization of nucleon wavefunctions,
 since it cancels in the ratio.
 We have
\begin{equation} \label{ratio}
 \left( \frac{\delta G_M}{G_M}, \frac{\delta\CG_3}{G_M} \right) =
   -\frac{3\alpha}{\alpha_s}
    \frac{\<\phi(y_i)|(T_{\delta G_M},T_{\delta\CG_3})|\phi(x_i)\>}
    {\<\phi(y_i)|T_{G_M}|\phi(x_i)\>}
\end{equation}
where $T_{G_M}$, $T_{\delta G_M}$ and $T_{\delta\CG_3}$ are the expressions,
sandwiched between wavefunctions in Eqs.~(\ref{**},\ref{delta}).

 The obtained TPE amplitudes are free from infra-red (IR) divergence.
 This becomes clear if we recall that the IR-divergent terms
 are proportional to Born amplitude. Thus $\delta\CG_3$ is IR-finite
 (it vanishes in Born approximation), and
\begin{equation}
 \delta G_M^{(\rm IR)} \sim \alpha G_M \ln \la^2
\end{equation}
 where $\la$ is infinitesimal photon mass.
 The magnetic form factor [Eq.~(\ref{**})] is a quantity of order
 $O(\alpha_s^2)$. On the other hand, leading-order contribution to TPE
 is $O(\alpha_s)$ and therefore IR divergence should only
 appear as a {\it subleading} effect, in the next order in $\alpha_s$.

 Another interesting point pertains to photons' virtualities.
 In all diagrams they are both of order $Q^2$, e.g. in the first diagram
 $q_1^2 = -x_2 y_2 Q^2$ and $q_2^2 = -(x_1+x_3)(y_1+y_3) Q^2$.
 We may conclude that the leading contribution to the amplitude
 at high $Q^2$ comes from the region where both photons are hard,
 $q_1^2 \sim q_2^2 \sim Q^2$.

\subsection{Wavefunctions}

Before turning to numerical calculations, we must specify
a model for wavefunctions.
The requirement for total spin and isospin to be 1/2
together with Pauli principle fix the following form
of the quark distribution amplitude (for the proton of positive helicity):
\begin{equation}
 |\phi(x_i)\> = \frac{f_N}{\sqrt{6}} \phi_1(x_1,x_2,x_3)
 \left( | u_\up u_\down d_\up \> - | u_\up d_\down u_\up \> \right) + {\rm perm.}
\end{equation}
where "perm." means sum over all quarks permutations and
$f_N$ is overall normalization constant not needed for our calculation.
The neutron wavefunction is obtained by interchange $d \leftrightarrow u$.
As we can see, the distribution amplitude is completely determined
by the function $\phi_1$.

In general, the distribution amplitude and thus $\phi_1$
depend logarithmically on $Q^2$. Namely, we have
\begin{equation}
 \phi_1(x_i,Q^2) = x_1 x_2 x_3 \sum_k \left[ \alpha_s (Q^2) \right]^{\gamma_k}
 B_k P_k(x_1,x_3)
\end{equation}
where $P_k$ are Appell polynomials ($P_1 = 1$, $P_2 = x_1-x_3$, etc.)
and $\gamma_k$ are corresponding anomalous dimensions \cite{BL}.
Thus in the formal limit $Q^2 \to \infty$ the term with lowest $\gamma_k$,
which is $P_1$, dominates and $\phi_1 \to \phi_{as} = x_1 x_2 x_3$.
This asymptotical wavefunction, however, leads to predictions inconsistent
with experiment. In particular, it yields zero proton and positive neutron
magnetic form factors.
Thus at present $Q^2$ the distribution amplitude should be considerably
different from its asymptotic form \cite{CZ}.
Since the evolution with $Q^2$ is very slow ($\gamma_k \ll 1$),
the same wavefunction can be employed for all currently accessible $Q^2$
with a reasonable accuracy.

Various forms of distribution amplitude were proposed in the literature.
We done the calculations with the following amplitudes: CZ~\cite{CZ},
KS~\cite{KS}, COZ~\cite{COZ}, GS~\cite{GS} and Het~\cite{Het}.
The CZ and KS amplitudes give practically
the same results as the COZ amplitude, thus they are not considered further.
 
\section{Numerical results} \label{sec:res}

\begin{figure}[t]
\hfill
\parbox{0.46\textwidth}
 { \includegraphics[width=0.45\textwidth]{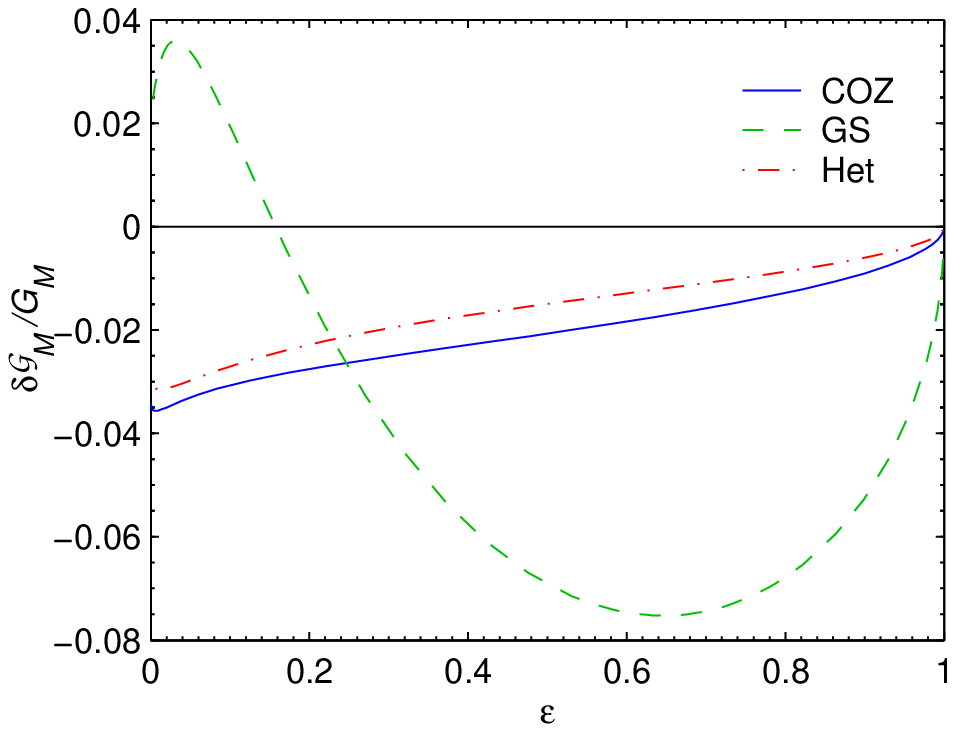} %{dCG_Mp-e.eps}
   \caption{TPE amplitude $\delta\CG_M/G_M$ vs. $\ve$ % for proton
     at $Q^2 = 10\GeV^2$} \label{fig:eps}
 }
\hfill
\parbox{0.46\textwidth}
 { \includegraphics[width=0.45\textwidth]{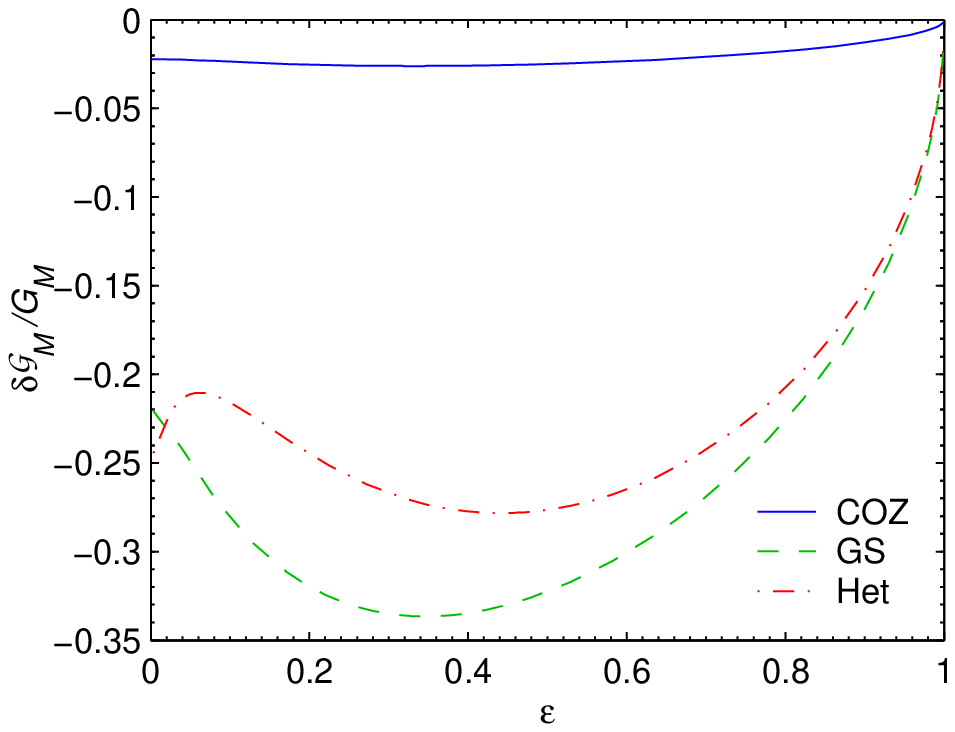} %{dCG_Mn-e.eps}
   \caption{TPE amplitude $\delta\CG_M/G_M$ vs. $\ve$ for neutron
     at $Q^2 = 5\GeV^2$} \label{fig:eps-neutron}
 }
\end{figure}
\begin{figure} \centering
 \includegraphics[width=0.45\textwidth]{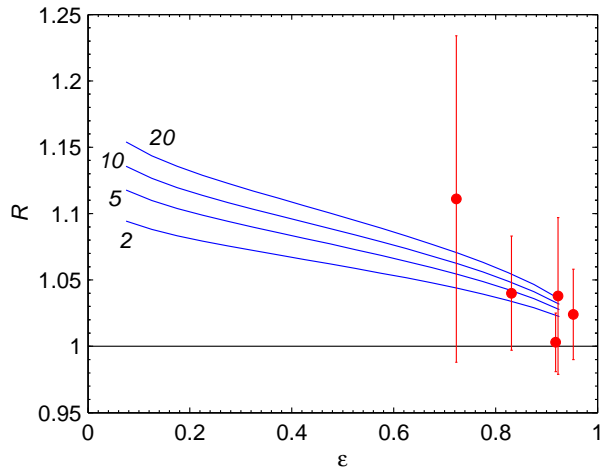}
  \caption{Positron/electron cross section ratio for $Q^2 = 2,5,10$ and
  $20 \GeV^2$ (shown near the curves).
  Data are from Ref.~\protect\cite{Mar} at $1.5 < Q^2 < 5 \GeV^2$.}
  \label{e+/e-}
\end{figure}

There are two independent kinematical variables in any elastic process.
For $eN$ scattering $Q^2$ and $\ve$ [Eq.~(\ref{epsilon})] are generally used.
The $\ve$-dependence of the obtained TPE amplitude $\delta\CG_M$
%, determined by the last term in Eq.~(\ref{ratio}),
is shown in Fig.~\ref{fig:eps}.
It turns out to be universal for all $Q^2$ (except for slow
logarithmic evolution, which we neglect here).
We see that the amplitude $\delta\CG_M$, calculated with COZ and Het
wavefunctions, is very close to the linear function of $\ve$.
Slight deviations from linearity are present near the endpoints $\ve=0$,
$\ve=1$ only. In contrast, GS wavefunction yields much larger
and highly nonlinear TPE amplitude.
In light of this it is worth noting that linear $\ve$-dependence of
$\delta\CG_M$ is necessary and sufficient for Rosenbluth plots
to remain linear even under the influence of TPE \cite{OurPheno}.
Since careful analysis of experimental data do not reveal any nonlinearity
in Rosenbluth plots \cite{Nonlin}, we conclude that the experiment disfavors
GS wavefunction.

For the neutron target, both GS and Het wavefunctions yield nonlinear
and anomalously huge TPE corrections, up to 25\% (Fig.~\ref{fig:eps-neutron}).
Taking into account the smallness of neutron electric form factor,
these corrections would manifest as severe nonlinearities
of Rosenbluth plots, that is, strong $\ve$-dependence
of the elastic cross section.
Though such cross section behaviour seems unlikely, the high-$Q^2$
neutron form factor data are too poor to draw a final conclusion.
Further experimental study of electron-neutron elastic scattering
at high $Q^2$ and different $\ve$ can show definitely whether
the nucleon is described by Het or by COZ wavefunction.
For the present moment we take the COZ wavefunction
as the most plausible.

The amplitude $\delta\CG_3$, which determines the correction
to longitudinal recoil polarization [Eq.~(\ref{P_l})],
is small ($<1\%$) for both proton and neutron, and
unfortunately lies below the precision of today's experiments.

The positron/electron cross section ratio is shown in Fig.~\ref{e+/e-}.
The calculation is done with COZ wavefunction at $Q^2 = 2$, 5, 10 and $20\GeV^2$.
The experimental data in the range $1.5 < Q^2 < 5 \GeV^2$ from Ref.~\cite{Mar}
are also shown. Though the data points are well near the curves,
the errors are very large. More precise data would be helpful,
preferably in the low-$\ve$ region, where the predicted ratio is higher.

\begin{figure}[t]
\centering
\includegraphics[width=0.45\textwidth]{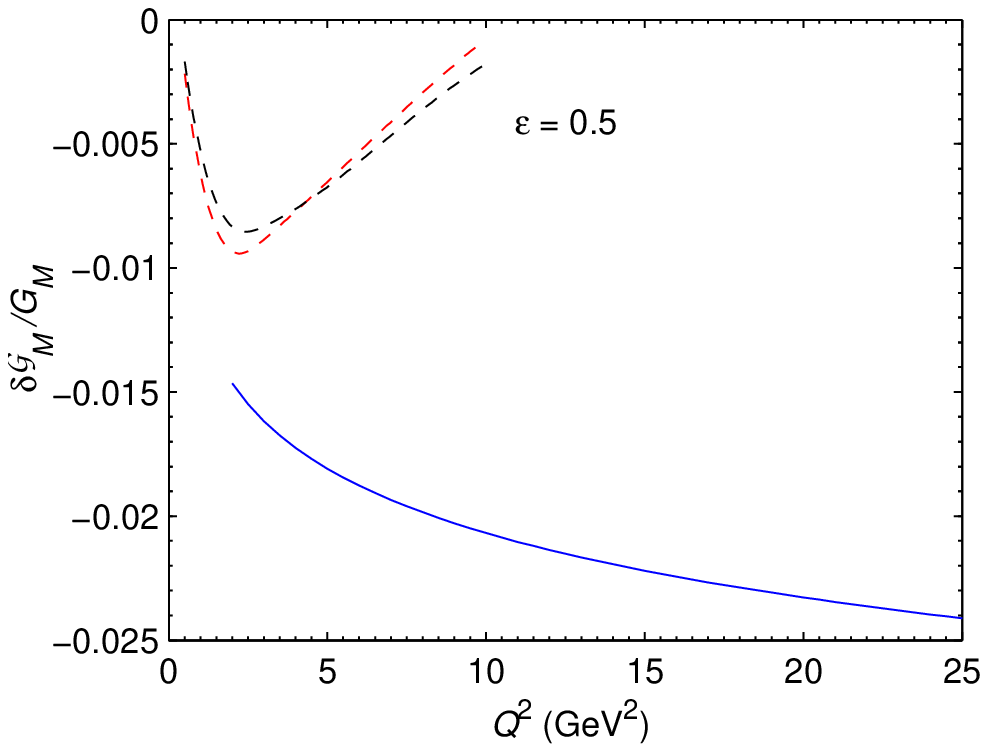}
\hfill
\includegraphics[width=0.45\textwidth]{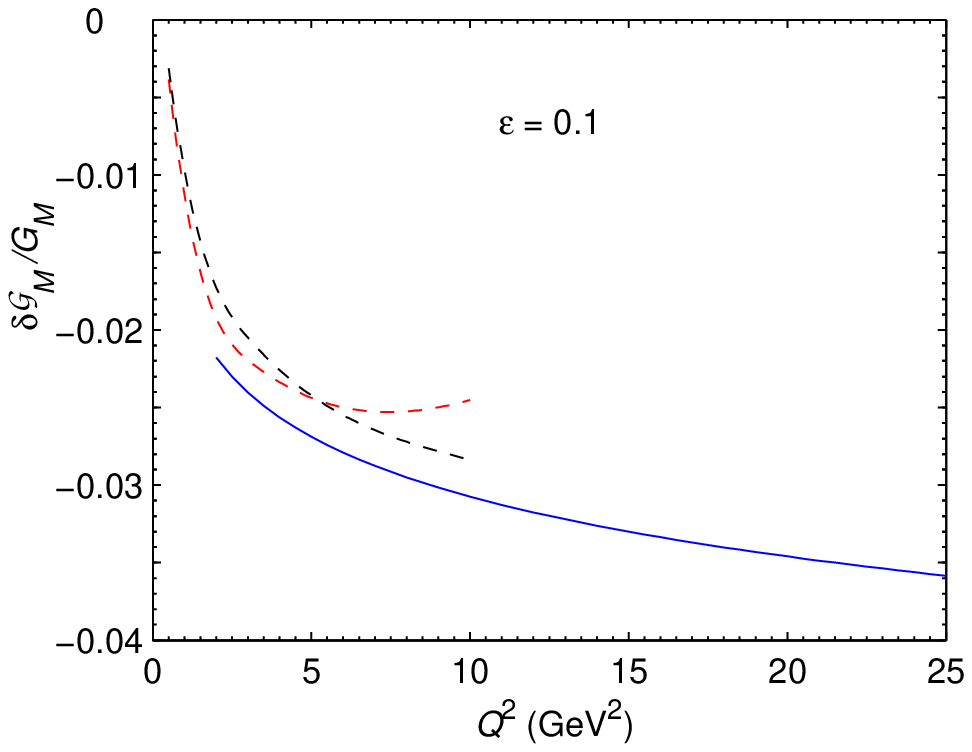}
\\
 \caption{TPE amplitude $\delta\CG_M$ vs. $Q^2$ at $\ve=0.5$ (left)
 and $\ve = 0.1$ (right). Dashed curves show hadronic calculations,
 with form factors parameterizations: dipole (red) and
 Ref.~\protect\cite{JLabFit} (black). } \label{fig:q2}
\end{figure}

The $Q^2$ dependence of "normalized" TPE amplitudes at fixed $\ve$
is completely determined by evolution of strong coupling constant $\alpha_s$.
We have used simple parameterization
\begin{equation}
 \alpha_s = \frac{4\pi}{\beta \ln ( Q^2/\La^2 )}
\end{equation}
with $\La = 0.2 \GeV$. The resulting shape of TPE amplitude $\delta\CG_M$
for proton, calculated with COZ wavefunction, is plotted in Fig.~\ref{fig:q2}.
At $Q^2\approx 30\GeV^2$, which is today the maximal $Q^2$ ever investigated,
the relative value of TPE amplitude reaches 3.5\%, which corresponds
to cross section correction of about 7\%.
Such a correction is however smaller than the errors of available data.
On the other hand, TPE can be seen in recently proposed high-$Q^2$
JLab experiment \cite{highQ2-p}, where the estimated errors are at 1\% level.

The results of "hadronic" calculation \cite{OurBox,OurDisp} are also shown
in Fig.~\ref{fig:q2} for comparison.
% We expected this curve to connect smoothly with pQCD prediction at higher $Q^2$
Probably, the amplitude undergoes some gradual transition from
this curve at lower $Q^2$ to pQCD prediction at higher $Q^2$
(recall that $\ve$-dependence in both cases is the same,
approximately linear with positive slope).
The figure suggests that a reasonable interpolation is possible between
the "hadronic" result for $Q^2$ below $\sim 3 \GeV^2$ and pQCD result
above this value.
But we also see a strong disagreement of these two curves at higher $Q^2$.
The most likely reason for such behaviour is that
the "hadronic" approach, % of Refs.~\cite{OurBox,BMTRes},
 i.e. saturation of the intermediate
 hadronic states by the bare nucleon and the lowest resonances,
 is inadequate at high $Q^2$. The multi-particle intermediate states
 yield a substantial part of the amplitude.

\section{Conclusions} \label{sec:concl}
We have considered TPE for the elastic electron-nucleon scattering
in the framework of pQCD. The calculations are done in the leading
order with several model wavefunctions.
For the proton target and wavefunctions based on QCD sum rules
(CZ \cite{CZ}, KS \cite{KS} and COZ \cite{COZ}),
the TPE amplitude $\delta\CG_M$,
which determines cross section correction, has linear $\ve$-dependence.
Its value is of order $\alpha/\alpha_s$, grows logarithmically with $Q^2$
and at $Q^2 = 30\GeV^2$ reaches 3.5\% of Born amplitude.
At lower $Q^2$ a smooth connection is possible with previous "hadronic"
calculations, in which TPE amplitudes were calculated
taking into account just the nucleon intermediate state \cite{OurBox}.
On the other hand, at high $Q^2$ the results of these two methods
are very different, which implies that "hadronic" approach becomes
inadequate at $Q^2 \gtrsim 3 \GeV^2$.

The size and $\ve$-dependence of TPE amplitudes
are sensitive to the choice of nucleon wavefunction
(quark distribution amplitude).
At the same time, they are directly measurable: $\delta\CG_M/G_M$ via
cross section or positron/electron cross section ratio and
$\delta\CG_3/G_M$ --- via longitudinal recoil polarization.
Thus an accurate measurement of TPE observables opens a new efficient way
to study quark distribution amplitude in the nucleon.
For example, the existing experimental data already rule out
GS wavefunction (Ref.~\cite{GS}).
Since TPE amplitudes have non-trivial $\ve$-dependence,
they potentially provide much more information,
than just nucleon form factors. % or exclusive charmonium decay widths.
Thus TPE turns from the correction to form factor measurements
into an independent tool for studying nucleon structure.

\begin{acknowledgements}
This work was supported by Program of Fundamental Research of the
Department of Physics and Astronomy of National Academy of Sciences of
Ukraine.
\end{acknowledgements}

\end{document}